\documentclass[twocolumn]{aastex631}
\newcommand{\lensedsn}{SN\,H$0$pe}

\usepackage{todonotes}
\usepackage{multirow}
\usepackage{amsmath}

\usepackage{xcolor}
\revised{March 26, 2026}

\begin{document}

\title{Testing Lens Models of PLCK G165.7+67.0 Using Lensed SN H0pe}
\author[0009-0008-1965-9012]{Aadya~Agrawal}
\affiliation{Department of Astronomy, University of Illinois Urbana-Champaign, 1002 West Green Street, Urbana, IL 61801, USA}
\affiliation{NSF-Simons AI for the Sky (SkAI) Institute, 875 N. Michigan Ave., Suite 3500, Chicago, IL 60611, USA}

\author[0000-0002-2361-7201]{J.~D.~R.~Pierel}
\affiliation{Space Telescope Science Institute, 3700 San Martin Drive, Baltimore, MD 21218, USA}

\author[0000-0001-6022-0484]{Gautham Narayan}
\affiliation{Department of Astronomy, University of Illinois Urbana-Champaign, 1002 West Green Street, Urbana, IL 61801, USA}
\affiliation{NSF-Simons AI for the Sky (SkAI) Institute, 875 N. Michigan Ave., Suite 3500, Chicago, IL 60611, USA}

\author[0000-0003-1625-8009]{B. L. Frye}
\affiliation{Department of Astronomy/Steward Observatory, University of Arizona, 933 N. Cherry Avenue, Tucson, AZ 85721, USA}

\author[0000-0001-9065-3926]{Jose M. Diego}
\affiliation{Instituto de Física de Cantabria (CSIC-UC). Avda. Los Castros s/n. 39005 Santander, Spain}

\author[0000-0003-3418-2482]{Nikhil Garuda}
\affiliation{Department of Astronomy, The University of Texas at Austin, 2515 Speedway Boulevard, Austin, TX 78712, USA}

\author[0000-0002-6741-983X]{Matthew~Grayling}
\affiliation{Institute of Astronomy and Kavli Institute for Cosmology, University of Cambridge, Madingley Road, Cambridge, CB3 0HA, UK}

\author[0000-0002-6610-2048]{Anton M. Koekemoer}
\affiliation{Space Telescope Science Institute, 3700 San Martin Drive, Baltimore, MD 21218, USA}

\author[0000-0001-9846-4417]{Kaisey~S.~Mandel}
\affiliation{Institute of Astronomy and Kavli Institute for Cosmology, University of Cambridge, Madingley Road, Cambridge, CB3 0HA, UK}

\author[0000-0002-2282-8795]{M.~Pascale}
\affiliation{Department of Astronomy, University of California, 501 Campbell Hall \#3411, Berkeley, CA, 94720, USA}

\author[0000-0001-7610-5544]{David Vizgan}
\affiliation{Department of Astronomy, University of Illinois Urbana-Champaign, 1002 West Green Street, Urbana, IL 61801, USA}

\author[0000-0001-8156-6281]{Rogier A. Windhorst}
\affiliation{School of Earth and Space Exploration, Arizona State University, Tempe, AZ 85287-6004, USA} 




\begin{abstract}

Supernova H0pe is a multiply-imaged Type Ia supernova (SN~Ia) and the second lensed SN to yield a measurement of the Hubble constant by the time-delay cosmography method, finding $H_0 = 75.4^{+8.1}_{-5.5} \text{km s}^{-1} \text{Mpc}^{-1}$  \citep{Pascale_SNH0pe_2025}. We investigate the seven lens modeling approaches used to derive $H_0$, assessing their agreement with $\Lambda \text{CDM}$ constraints from SN~Ia surveys through a purely observational comparison.  {We test each lens model by combining its predicted magnifications with the observed time delays to reconstruct the intrinsic SN~Ia luminosity and corresponding distance modulus.} While photometrically derived magnifications yield distance moduli in line with $\Lambda \text{CDM}$ expectations, our comparison reveals that lens model predictions, even the most precise ones,  {consistently overestimate the magnification, with an offset $> 1$~mag}. This known bias, already appreciated by modeling teams, is independently confirmed through our analysis and highlights the value of lensed SNe as a tool to test model accuracy.  {If unaccounted for, such magnification biases can propagate into uncertainties in derived cosmological parameters, including $H_0$, a critical challenge for precision cosmology using strongly lensed transients.} 
\end{abstract}




\section{Introduction}\label{sec:intro}

\begin{figure*}[!ht]
    \centering
    \includegraphics[width=0.85\linewidth]{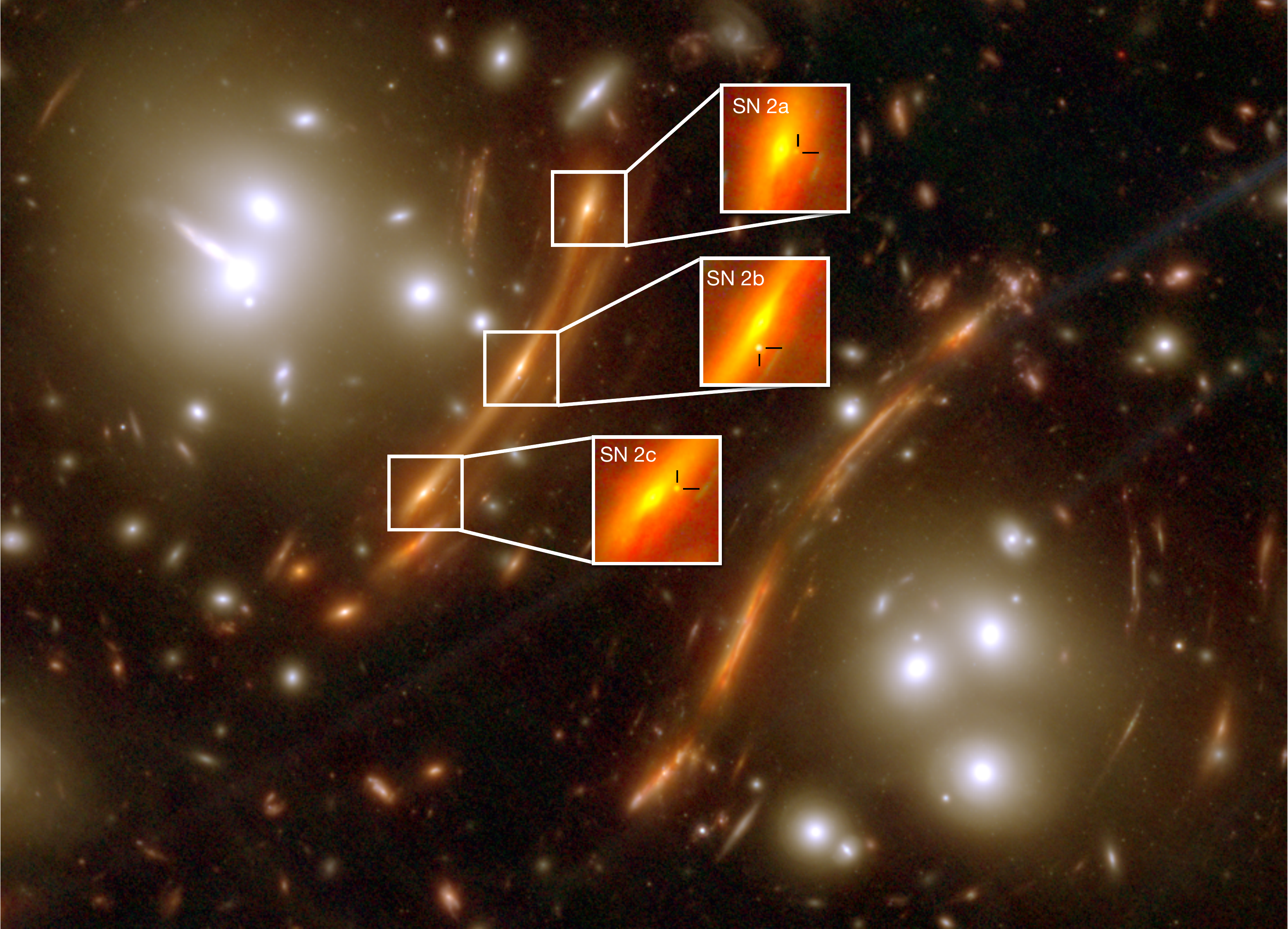}
    \caption{Mosaic of the central region of the PLCK G165.7+67.0 cluster field, covering a width of 1 arcminute, showing the full depth obtained in all 5 epochs for all 6 NIRCam filters F090W, F150W, F200W, F277W, F356W, and F444W (see Section \ref{sec:obs} for further details). The insets show a close up of each of the images of SN H0pe based on the JWST Epoch 1 data. Each of the insets has a partial crosshair that shows the location of the SN.}
    \label{fig:color_im}
\end{figure*}

Supernovae have been central to astronomical discovery for centuries, beginning with Tycho Brahe’s 1572 observation of a ``new star'' that overturned the prevailing notion of an immutable sky. Type Ia supernovae (SNe Ia), are thermonuclear explosions of CO white dwarfs and function as ‘standardizable candles’, enabling astronomers to measure the distances to extragalactic sources \citep{standard_candles_1992, Tripp_1998, Guy_2007_SALT,SN_cosmology_review, Pantheon_2022}. SNe~Ia also serve as probes to measure the Hubble constant ($H_0$), a cosmological parameter that quantifies the local rate of expansion of the universe \citep{Riess_1998, Perlmutter_1999}. Measuring $H_0$ provides a critical test of the standard $\Lambda \text{CDM}$ cosmological model \citep{LCDM_review_2018, lcdm_H0_review_2022}. 

The ``local distance-ladder'' method, using Cepheids and SNe~Ia, is one of the primary methods used to measure $H_0$ \citep{Riess_2024_review}. The `Supernova $H_0$ for the Equation of State' (SH0ES) team \citep{Riess_2024} measured a value of $H_0 = 72.6 \pm 2.0$ km s$^{-1}$ Mpc$^{-1}$. This value differs from the value of $H_0 = 67.4 \pm 0.5$ km s$^{-1}$ Mpc$^{-1}$ found by \citet{Planck_2020} using measurements from the Cosmic Microwave Background (CMB) in conjunction with $\Lambda \text{CDM}$. This `Hubble Tension' at the $5\sigma$ level \citep{Verde_2019, Riess_2016}, represents one of the most significant issues in cosmology today. Many solutions have been proposed to resolve the discrepancy \citep{Weinberg_2013}, including using new probes such as redshift-space distortions \citep{RSD_tension}, standard sirens \citep{standard_sirens_1, standard_sirens_aas} and precision gravity tests\citep{MOND}. 

Assuming that both the SNe~Ia and CMB measurements are accurate, several recent studies have explored modifications to the standard cosmological model--i.e., Early Dark Energy \citep[i.e.,][]{kamionkowski_early_de_2023}, Late Dark Energy \citep[i.e.,][]{late_de}, and Dark Energy models with additional degrees of freedom \citep[i.e.,][]{Di_Valentino_2021}--in an effort to reconcile the differing measurements. Each of these proposed scenarios has significant implications for processes such as star formation, galaxy evolution, and large-scale structure, requiring thorough scrutiny. Another possibility for the source of the tension lies in a systematic error in one of the many probes. Larger surveys and more precise measurements are being taken and planned to determine the cause. 

One promising approach to addressing the Hubble tension is to leverage strongly gravitationally lensed supernovae (glSNe) for cosmological measurements and probes of $H_0$ \citep{Refsdal_1964, Suyu_24_TDC, Goobar_2025}. Strong gravitational lensing occurs when light from a background source bends around a massive foreground system, causing the light to take different paths to reach the observer and resulting in multiple images of the source \citep{Einstein_1936, Zwicky_lenses_1937, Zwicky_nebulae_1937}. When a SN explodes in a multiply-imaged host galaxy, the light from the SN follows paths of varying length and gravitational potential, resulting in both a geometric and a Shapiro (gravitational) time delay. Together, these effects produce an observable time delay between images, determined by the distribution of the gravitational potential of the foreground lens system \citep{Narayan_Bartelmann_1996, Oguri_2019}. This time delay depends on the angular diameter distances of the source, lens and observer relative to each other. As these distances are dependent on cosmology, measuring the time delays in turn can be used to probe cosmological parameters. When combined with a lens model of the foreground mass distribution, these time delays allow glSNe to serve as one-step probes of $H_0$ \citep{Refsdal_1964} in a method called Time Delay Cosmography. The lens model provides the difference in Fermat potential between the image positions, which (combined with the observed time delays) allows for the inference of the `time-delay distance,' and thus, $H_0$. Since this distance scales inversely with $H_0$, accurate modeling of the lens potential is essential to extract reliable cosmological constraints. 

While other transients can also be used for Time Delay Cosmography, SNe are particularly advantageous for this method \citep{Pierel_PhotometricTimedelay_2024a}. As the SN flux fades, images of the host without the SN flux can be taken to get more accurate photometry without host contamination, providing a more precise $H_0$ value \citep{Ding_2021}. Additionally, the non-repeating and non-stochastic light curve of glSNe can be leveraged to break degeneracies in the lens models which is difficult for other lensed objects that permit time delay cosmography e.g., quasars \citep{Birrer_2024_review, Arendse_2024}. A glSN~Ia, in particular, provides even more precision as there exist well-understood models of SN~Ia evolution \citep{Guy_2007_SALT,Kenworthy_2021_SALT3,SALT3NIR_2022,Mandel_2022, Ward_2023_SNIa, grayling_24_mnras}, which makes determining time-delays simpler. Furthermore, a glSN~Ia enables the measurement of the absolute magnifications of its multiple images—an observable that directly informs the lensing geometry and the inferred value of the Hubble constant. As demonstrated by \citet{Liu_Oguri_2025}, combining time-delay measurements with absolute magnification constraints significantly improves the precision of $H_0$ inference by breaking degeneracies in the lens model. This joint analysis not only tightens the posterior on $H_0$ but also helps mitigate biases that arise from uncertainties in the lens potential, especially in regions with limited imaging constraints. 

A handful of glSNe have been discovered in recent years. 
SN Refsdal, the first multiply-imaged SN, enabled a measurement of the time delay, and from it, a value for $ H_0 = 66.6^{+ 4.1}_{-3.3}$ km s$^{-1}$ Mpc$^{-1}$  by the method of time-delay cosmography \citep{Kelly_2023_a, Kelly_2023_b}. Subsequently discovered glSNe such as SN Requiem \citep{Rodney_2021a}, AT2022riv \citep{AT2022riv} and C22 \citep{Chen_2022} were either detected too late to observe multiple images or found in archival imaging, limiting opportunities for follow-up. However, SN Requiem may offer another opportunity, with a predicted reappearance expected around ~2027 \citep{Suyu_2025_Encore_Requiem}. SN Zwicky \citep{Goobar_2023_zwicky, Pierel_2023_zwicky, Larison_2025_zwicky} and iPTF16geu \citep{Goobar_2017}, two glSNe~Ia identified in galaxy-scale lenses, were ultimately too compact to provide meaningful constraints on $H_0$. \lensedsn~was the first glSN~Ia with sufficient data and time delays long enough to facilitate the  measurement of $H_0$ \citep{Frye_2023, Frye_2024}.  {This single SN appeared in three positions in the image plane as result of strong lensing. This triply-imaged SN enabled the measurement of the two time delays by a photometric approach \citep{Pierel_PhotometricTimedelay_2024a}, and a rare {\it spectroscopic} one \citep{chen_SN_2024}.} Owing to the nature of this SN as a standard candle, absolute magnifications were also measured for all three SN appearances as a part of these photometric and spectroscopic programs, and this set of five observables was inducted into the $H_0$ inference \citep{Pascale_SNH0pe_2025}. Seven lens models were constructed from seven different approaches, and the results of all groups were double-blinded by strictly enforced protocols. On scaling the lens-predicted values for the observables to the measured ones, a value of $H_0 = 75.4^{+8.1}_{-5.5} \text{km s}^{-1} \text{Mpc}^{-1}$ was inferred \citep{Pascale_SNH0pe_2025}. This result highlights the potential for using such systems to address the Hubble Tension. The most recent glSN~Ia to be observed is SN Encore, incredibly found in the same galaxy as SN Requiem, with the probability of $\lesssim 3\%$ for detecting two SNe-Ia in this system in the last decade \citep{Pierel_Encore_2024}. SN Encore found an measurement of $H_0 = 66.9^{+11.2}_{-8.1} \text{km s}^{-1} \text{Mpc}^{-1}$ in a double blind analysis \citep{Pierel_2025_Encore, Suyu_2025_Encore_Requiem}. 

Lensed SNe~Ia also present a unique opportunity to test existing lens models \citep{Rodney_testing, Patel_2014}. Among different lens modeling methods, one of the most common ones is based on the assumption that light traces mass or `LTM' \citep{Zitrin_2009}. This means that the galaxies in a cluster act as tracers of the dark matter distribution. This method is well-tested and has proven to result in robust lens models that can replicate observations well. However, when predicting the reappearance of SN Refsdal, \citet{Zitrin_2021} discovered a minor but pivotal numerical artifact that had perviously gone unnoticed in the Time Delay calculation. While this particular issue could also have been revealed through other checks, it illustrates a broader benefit: lensed transients provide a direct way to evaluate the underlying methodology and assumptions of lens models. Despite the low probability of source, lens, and observer alignments, the \textit{James Webb Space Telescope (JWST)} has enabled the observation of many multiply-imaged systems. In the coming years, facilities like the \textit{Vera C. Rubin Observatory}, \textit{Nancy Grace Roman Telescope}, and others are expected to observe $\sim 100$ glSNe \citep{LSST_Roman_rates, Arendse_2024, Pierel_Roman_rates}. The dramatic increase in the number of observed glSNe systems necessitates robust lens models. 

\begin{figure}[t]
    \centering
    \includegraphics[trim= 0.5cm 1.5cm 0.5cm 1.5cm, clip, width=\linewidth]{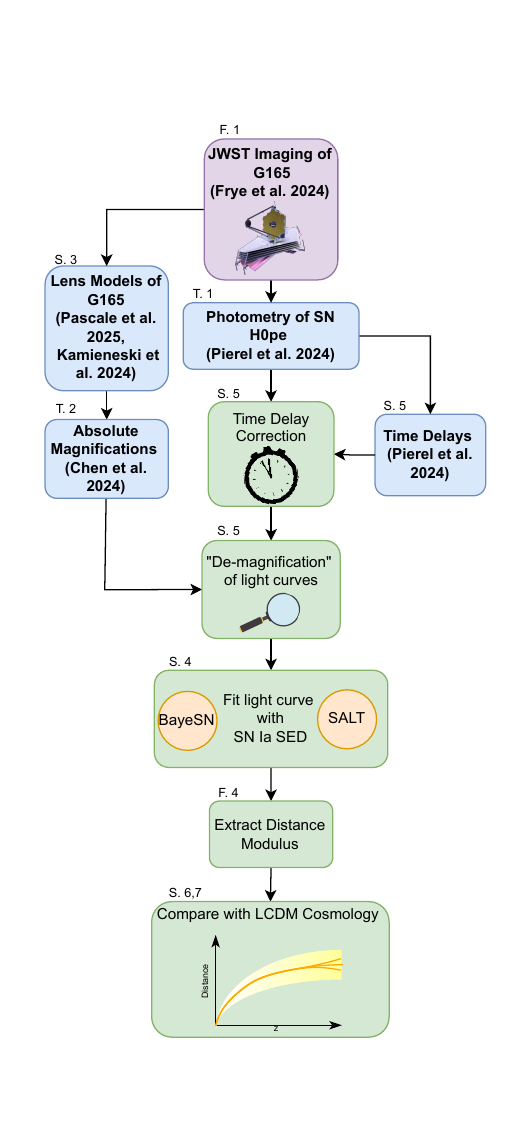}
    \caption{A schematic showing the method used in this analysis. Each step has an associated section (S.), table (T.) or figure (F.) in this paper, that provides more details about the work.}
    \label{fig:schematic}
    \vspace{-2mm}
\end{figure}

 {We set out to quantify the agreement between the light curves predicted by lens models of PLCK G165.7+67.0 drawn from \citet{Pascale_SNH0pe_2025} and those of a typical SN Ia following assumed cosmological models using observations of \lensedsn. Figure \ref{fig:color_im} shows the full-depth mosaic from combining all 5 epochs of the PLCK G165.7+67.0 cluster field, along with insets showing close-up images of \lensedsn~as observed in Epoch 1. We use the measured time-delays and model-predicted magnifications from \citet{Pascale_SNH0pe_2025} for each lens model to correct for lensing effects and reconstruct an `intrinsic' light curve for \lensedsn~by jointly analyzing all three images. This de-magnified intrinsic light curve is fit using established SN~Ia spectral energy distribution models to infer a distance modulus. By leveraging the standardizable nature of \lensedsn, we place this multiply imaged SN~Ia on the same distance scale as the cosmological SN~Ia sample, enabling the first direct comparison between a multiply imaged standard candle and the broader SN~Ia population. Our analysis rests on the assumption that, if the lens models accurately predict the magnification at the image positions, the inferred distance modulus should be consistent with that of a `typical' SN~Ia at $z=1.783$, as expected from cosmological parameters derived from previous SN~Ia datasets.} Figure \ref{fig:schematic} shows the workflow and data used for this analysis. The figure includes information on section, table or figure where the elements are described in this paper which is organized as follows. 

In Section \ref{sec:obs} we describe the observations and data used in this analysis. A brief description of the lens models taken from the \citet{Pascale_SNH0pe_2025} are given in Section \ref{sec:lens_models}. In Section \ref{sec: sed modeling}, we detail the methods used to conduct the analysis. This analysis relies heavily on SN~Ia SED models and thus, under Section \ref{sec: sed modeling}, Subsection \ref{subsec: sed models} describes each model and Subsection \ref{subsec: getting mu} describes the methods used to calculate the distance fitted parameters from each of the models. In Section \ref{sec:results} we state the results of the analysis which are further discussed in Section \ref{sec: concl}.  {Throughout this paper, $\mu$ is used to refer to luminosity distance measurements, and not magnification due to strong lensing.}

\begin{table*}[!ht]
    \caption{Photometry of \lensedsn~as given in \citet{Pierel_PhotometricTimedelay_2024a}}
    \centering
    \begin{tabular*}{\linewidth}{@{\extracolsep{\stretch{1}}}*{7}{c}}
    \toprule
    MJD & Filter & Exp Time (s) & $m_{2a}$ & $m_{2b}$ & $m_{2c}$ \\
    \hline
    60033 & F090W & 2490 & $30.55 \pm 0.47$ & $25.53 \pm 0.05$ & $ 28.71 \pm 0.11$ \\
    60033 & F115W & 2490 & $28.35 \pm 0.11$ & $24.96 \pm 0.06$ & $ 26.16 \pm 0.07$ \\
    60033 & F150W & 1889 & $26.50 \pm 0.08$ & $23.93 \pm 0.07$ & $ 24.95 \pm 0.07$ \\
    60033 & F200W & 2104 & $26.45 \pm 0.08$ & $24.21 \pm 0.08$ & $ 25.25 \pm 0.08$ \\
    60033 & F277W & 2104 & $25.72 \pm 0.14$ & $24.93 \pm 0.14$ & $ 25.58 \pm 0.14$ \\
    60057 & F090W & 1245 & $>29.71 $        & $26.44 \pm 0.07$ & $ 30.23 \pm 0.74$ \\
    60057 & F150W & 858  & $27.06 \pm 0.10$ & $24.03 \pm 0.07$ & $ 25.45 \pm 0.08$ \\
    60057 & F200W & 1760 & $26.82 \pm 0.09$ & $24.35 \pm 0.08$ & $ 25.06 \pm 0.08$ \\
    60057 & F277W & 1760 & $26.07 \pm 0.15$ & $25.20 \pm 0.14$ & $ 25.37 \pm 0.14$ \\
    60074 & F090W & 1417 & $>29.48 $        & $27.21 \pm 0.09$ & $ 30.45 \pm 0.81$ \\
    60074 & F150W & 1245 & $27.37 \pm 0.12$ & $24.42 \pm 0.07$ & $ 25.94 \pm 0.08$ \\
    60074 & F200W & 1760 & $26.74 \pm 0.09$ & $24.81 \pm 0.08$ & $ 25.24 \pm 0.08$ \\
    60074 & F277W & 1760 & $26.28 \pm 0.15$ & $25.21 \pm 0.14$ & $ 25.18 \pm 0.14$ \\
    \hline
    \end{tabular*}
    \label{tab:phot_table}
\end{table*}

\section{Data and Observations} \label{sec:obs}

This Section summarizes: (1) the discovery and observations of \lensedsn~and (2) the procedures used to extract photometry for the three appearances \lensedsn.

\subsection{Summary of Observations}
\lensedsn~was discovered in JWST imaging of the PLCK G165.7+67.0 cluster field, taken on 30 March 2023 (MJD 60033) in eight NIRCam filters as part of the Prime Extragalactic Areas for Reionization and Lensing Science (PEARLS) program (PID 1176, PI: R~Windhorst). For details on the reduction and analysis of this dataset, we refer the reader to \citet{Windhorst_2023} and \citet{Frye_2023, Frye_2024}. In the NIRCam F150W image, three transient sources were identified by comparing to archival F160W Hubble Space Telescope (HST) imaging of the same field, acquired on 30 May 2016 (MJD 57538) under an unrelated program (GO-14223, PI: B.~Frye; \citet{Frye_2023}). The similarity in wavelength coverage and throughput between the two filters enabled this comparison.  {The detection of the three sources in the JWST data, the relatively high brightness of the brightest lensed source at approximately 24.3 (AB) mag, and the changing parity of the source-host galaxy pair on each crossing of the Einstein ring, pointed to a lensed SN. A SN estimator applied to the initial three light curve points yielded a 95\% probability that the event was a Type Ia supernova} \citep{Frye_2023}. This led to a JWST DDT proposal to monitor the SN by photometry and spectroscopy (PID 4446, PI:  B. Frye), and two ground-based proposals to obtain redshifts of the SN host galaxy using the Large Binocular Telescope (LBT; PI: B. Frye), and the MMT (B. Frye). The LBT observations enabled the measurement of the spectroscopic redshift of the SN host galaxy to be $z = 1.783 \pm 0.002$ \citep{Polletta_2023}, consistent with a prior photometric redshift estimate \citep{Pascale_2022}.

The JWST follow-up observations provided two additional epochs of NIRCam imaging in six filters (F090W, F150W, F200W, F277W, F356W, and F444W), along with NIRSpec Prism, G140M, and G235M spectroscopy of all three SN appearances and two images of the SN host galaxy. The NIRSpec spectroscopy enabled a redshift measurement for the third image of the host galaxy and refined redshift estimates for Arc 2a and Arc 2c to $z = 1.7833 \pm 0.0010$ and $ z = 1.7834 \pm 0.0005$, respectively \citep{chen_SN_2024, Frye_2024}. The follow-up imaging was taken on 22 April 2023 (MJD 60056) and 9 May 2023 (MJD 60073), resulting in three observing epochs spaced by approximately one rest-frame week, or about three observer-frame weeks. Two additional observing epochs were obtained as part of a JWST Cycle 3 program, which provides the first template image of Arc 2 (PID 4744, PIs: B.~Frye and J.~Pierel). 

\subsection{Photometry}
The steps taken to reduce the data set and perform the photometry are described in detail by \citet{Windhorst_2023} and \citet{Pierel_PhotometricTimedelay_2024a}. Briefly, the observations of \lensedsn~were calibrated using the STScI JWST Pipeline\footnote{\url{https://github.com/spacetelescope/jwst}} (Version 1.12.5; \citet{JWSTCalibrationPipeline_2024}) and were available as multiple data products. The products relevant for the photometric measurements were the level 2 ``CAL" images, which are individual exposures that have been calibrated with the pipeline, bias-subtracted, dark-subtracted, flat-fielded, but have not been corrected for geometric distortion and the level 3 drizzled \citep{multidrizzle_2003} images that combine the individual exposures \citep[common when analyzing dithered exposures; see][]{Windhorst_2023}\footnote{This data was obtained from the Barbara A. Mikulski Archive for Space Telescopes (MAST) at the STScI (accessible via \doi{10.17909/zk1p-2q51})}. Measuring the position and brightness of a SN is understood to yield the best results when done on the CAL images as this processing preserves the PSF structure better than the drizzled images \citep{Rigby_JWST_2023}, but the host galaxy brightness relative to the SN made this impractical for longer wavelengths. Photometry was therefore performed on the F150W level 2 images to achieve a baseline for comparison, but then a drizzled image PSF fitting routine was used to measure the final photometry, while the F150W CAL image photometry ensured accuracy. 

Another consideration when measuring SN photometry is separating the host flux from the SN flux. Due to the lack of a template image at the time, a lens model was used to create a host surface brightness profile. Subtracting this profile from the observed flux isolated the SN flux. We adopt the photometry reported in Table 2 of \citet{Pierel_PhotometricTimedelay_2024a}, reproduced here in Table \ref{tab:phot_table}. The photometry is reported in AB magnitudes and the upper limits for the F090W filter for two of the epochs are $3~\sigma$. A complete reanalysis that includes the fifth observing epoch (shown in Figure \ref{fig:color_im}) will appear in a future paper (Agrawal et al. 2026, \textit{in prep.})

\section{Lens Models}\label{sec:lens_models}

The seven lens models analyzed in this paper are described in detail by previous works \citep{Frye_2024, Kamieneski_2024, Pascale_SNH0pe_2025}. Briefly, there were seven lens modeling teams that analyzed the cluster and obtained the model-predicted time delays and absolute magnifications. All seven lens modeling subgroups agreed in advance to use a consistent set of 21 image systems (including five with spectroscopic redshifts). They also settled on the positions, F200W brightnesses, and morphological parameters of 161 galaxies present in the cluster, along with any known galaxy interlopers. The details are specified in \citet{Frye_2024}. Each subgroup conducted a double-blinded analysis without knowledge of the spectroscopic/photometric time delays or absolute magnifications or other lens modeling teams’ analyses that was strictly enforced. All models assumed a fiducial cosmology of $\Omega_m = 0.3$, $\Omega_\Lambda = 0.7$ and an $H_0 = 70 \text{ km s}^{-1} \text{Mpc}^{-1}$. 

A range of approaches from parametric to non-parametric  were used to build the seven lens models with one even combining strong and weak lensing. Parametric models use a superposition of analytic mass density profiles and assume they characterize the entire mass distribution, with a large cluster scale halo described by a large-scale profile and smaller halos for individual galaxies. In contrast, non-parametric models describe the mass distribution using a flexible grid with few assumptions about the dark matter profile of cluster halo. Semi-parametric models use a combination of the flexible grid-based mass distribution with analytic halo profiles.  
\begin{table*}[!ht]
    \caption{Predicted magnifications of \lensedsn~per image as given in and weights for each model as stated in Tables 1 and 2 in \citet{Pascale_SNH0pe_2025} }
    \centering
    \begin{tabular*}{\linewidth}{@{\extracolsep{\stretch{1}}}*{7}{c}}
    \toprule
    \# & Reference & $|\text{Mag}_a|$ & $|\text{Mag}_b|$ & $|\text{Mag}_{c}|$ & $\text{Weight}_{\text{TD-only}}$ \\
    \hline
    1 & \textsc{GLAFIC} \citep{Oguri_2019} & $8.02_{-0.57}^{+0.64}$ & $12.23_{-1.51}^{+1.30}$ & $9.32_{-0.87}^{+0.80}$ & 0.23 \\
    2 & \textsc{ Zitrin-Analytic} \citep{Zitrin_2009} & $11.25_{-0.90}^{+1.05}$ & $16.03_{-1.81}^{+1.77}$ & $14.48_{-1.65}^{+1.91}$ & 0.17\\
    3 & \textsc{LENSTOOL} \citep{kneib_2011} & $6.67_{-0.14}^{+0.11}$ & $9.82_{-0.31}^{+0.23}$ & $8.94_{-0.29}^{+0.21}$ & 0.09 \\
    4 & \textsc{ MARS} \citep{MARS_2022} & $6.82_{-0.44}^{+0.50}$ & $9.17_{-0.72}^{+0.80}$ & $7.55_{-0.70}^{+0.86}$ & 0.16 \\
    5 & \citep{chen_2020} & $6.52_{-0.22}^{+0.24}$   & $10.35_{-0.41}^{+0.46}$ & $6.68_{-0.22}^{+0.24}$ & 0.30 \\
    6 & \textsc{ WSLAP+} \citep{Diego_2005} & $28.37_{-7.15}^{+6.34}$ & $63.70_{-17.34}^{+16.70}$ & $36.04_{-9.52}^{+9.57}$ & 0.00 \\
    7 & \textsc{ Zitrin-LTM} \citep{Zitrin_2009} & $5.66_{-0.14}^{+0.15}$ & $9.77_{-0.51}^{+0.47}$ & $8.80_{-0.46}^{+0.67}$ & 0.03 \\
    \hline
    Photometry & \citet{Pierel_PhotometricTimedelay_2024a} & $4.43_{-1.60}^{+1.52}$ & $8.00_{-2.34}^{+3.42}$ & $6.43_{-1.13}^{+1.25}$ & - \\
    Spectroscopy & \citet{chen_SN_2024} & $10.93_{-5.16}^{+8.63}$ & $13.22_{-2.33}^{+7.49} $ & $7.14_{-1.66}^{+1.553}$ & - \\
    \hline
    \end{tabular*}
    
    \label{tab:models}
\end{table*}
The seven models are listed below with a few key points for each model summarized from the appendix of \citet{Pascale_SNH0pe_2025}.
\begin{itemize}
    \item{Model 1 is a parametric model built using \textsc{GLAFIC} \citep{glafic_oguri_2010, Oguri_2019, Oguri_2021} with five Navarro-Frank-White (NFW) \citep{NFW_1997} profiles to model the dark-matter halos. Cluster galaxies were modeled using pseudo-Jaffe ellipsoid profiles scaled by the F200W flux of each galaxy. The model increases computational efficiency by modifying resolution based on magnification gradients using an adaptive-mesh grid. The model also includes an external shear component for flexibility with a Markov Chain Monte Carlo method of $\sim 10^4$ steps used for optimization  {in the source plane.}}
    
    \item{ Model 2 is also a parametric approach built using a modified version of the method used in \citet{Zitrin_2015} (further details about the new version are given in \citet{Furtak_2023}). The cluster galaxies were modeled as double pseudo-isothermal elliptical mass-density distributions  {\footnote {This profile is mathematically equivalent to the pseudo-Jaffe profile used in Model 1 but differs in normalization across software implementations.}} (dPIE, \citet{eliasdottir_2007}) while the cluster-scale dark matter halo was described by a pseudo-isothermal elliptical mass- density distribution (PIEMD; \citet{kassiola_1993}). Flexibility was added by allowing independent scaling of four central galaxies. Circular profiles were assumed for all cluster galaxies. This model used MCMC for optimization and uncertainty calculation  {in the image plane.}} 
    
    \item{Model 3 is another parametric approach built using \textsc{LENSTOOL}\footnote{\url{ https://projets.lam.fr/projects/lenstool/wiki}} \citep{kneib_2011}, which uses the locations of the multiply-imaged systems to constrain the mass profiles. The member galaxies were assigned masses scaled to their F200W fluxes, relative to a characteristic $L^*$ galaxy ($m_{F200W} = 17.0$ AB mag) while the dominant merging components were modeled using two cluster-scale PIEMD halos. MCMC sampling with 10 chains of 1000 steps  {in the image plane} was used to fit the many free parameters of the model. Full details of the modeling approach are provided in \citet{Kamieneski_2024}.}
    
    \item{Model 4 was built using the \textsc{MARS} algorithm \citep{MARS_2022, MARS_2024}, a free-form, grid-based lens modeling method incorporating both strong lensing (SL) and weak lensing (WL), contrasting the parametric approaches taken in Models 1, 2 and 3. The \textsc{MARS} team provided two models to predict the magnifications and time delays: an SL-only model on a $100 \times 100$ grid ($90^" \times 90^"$), and a combined SL+WL model on a $400 \times 400$ grid ($360^" \times 360^"$). The SL-only model takes in a chi-squared minimization of multiple SL images of the source plane and a regularization term based on maximum cross entropy, which helps suppress noise and promotes a stable, quasi-unique mass reconstruction. The SL+WL model uses an additional term minimizing the difference between the observed and predicted reduced shear values.}
    
    \item{Model 5 is a semi-parametric approach and was originally built for a blind prediction of the reappearance of SN Refsdal \citep{chen_2020}. This model combines a non-parametric at the cluster-scale with analytic profiles at the individual galaxy-scale similar to parametric methods. This model uses symmetric analytic NFW profiles with the same scale radius and masses scaled using stellar flux (measured from JWST F090W imaging) to parameterize the galaxy dark matter halos. The model flexibly maps the cluster-scale dark matter distribution by applying smooth perturbations to the lensing potential without assuming symmetry.  {For each case of a source galaxy image that appears in multiple locations, those images should shrink to a single position on the source plane. This model uses the offsets in the source plane positions for optimization.}}
    
    \item{Model 6 is a hybrid model that was built using the \textsc{WSLAP+} software \citep{Diego_2005, Sendra_2014}. The model uses the light distribution of individual galaxies to trace the small-scale halos while describing the large-scale components using a grid of Gaussians. A uniform grid was recursively built and an adaptive grid was derived from it. This adaptive grid was then used to generate different models of the system and the time delay and magnifications. Limited constraints in certain regions of the cluster resulted in a large range of predicted magnifications. A system of two linear equations per lensing constraint are used to optimize the model. The model predicted that image C of \lensedsn~would be the last to arrive (inconsistent with observations) and hence did not have any weight in the $H_0$ inference done in \citet{Pascale_SNH0pe_2025}.}
    
    \item{ Model 7 is semi-parametric and uses a modified version of Zitrin-LTM  \citep{Zitrin_2009, Zitrin_2015, Broadhurst_2005} originally developed for time-delay cosmography of SN Refsdal. It uses the F200W brightness of the cluster-member galaxies as weights when estimating the stellar mass with a power-law mass-surface-density profile. These profiles are smoothed with a Gaussian kernel to approximate the cluster dark matter distribution. Systematics are accounted for by varying an external shear component. Individual galaxies like the BCG are fitted freely. Spectroscopic redshifts are used where available and photometric redshifts are input as best guesses for the remaining systems. An MCMC method is used to infer the model parameters in the image plane.}
\end{itemize}

 {Table \ref{tab:models} lists the predicted magnifications for each image of the SN for each model (extracted from \citet{Pascale_SNH0pe_2025}, their Table 1). In \citet{Pascale_SNH0pe_2025}, the value of $H_0$ is inferred from the time delays using a Bayesian framework. Each of the lens model predicted time delays assumed a fiducial $H_0 = 70 \text{km s}^{-1} \text{Mpc}^{-1}$. The fiducial time delays are scaled as a function of $H_0$ to match the observed photometric time delays, as shown in Equation \ref{eq:h0_rescale}.}

\begin{equation} \label{eq:h0_rescale}
    \Delta t_{i,j}^{pred} (H_0) = \Delta t_{i,j}^{fid} \times \frac{70 \text{ km s}^{-1}\text{Mpc}^{-1}}{H_0}
\end{equation}

 {The set of observables $\{\Delta t_{a,b},\, \Delta t_{b,c}\}$ each have measurements from the light curve and predictions from each lens model $M_\ell$ assuming a fiducial $H_0$; Equation~\ref{eq:h0_rescale} is used to rescale the fiducial time delays until the predictions match the LC measurements. The LC measurements themselves are obtained by fitting the light curves of the multiple images, marginalising over SN-model-specific nuisance parameters (e.g.\ light-curve shape) to obtain the posterior on the lensing observables. Analogously, each lens model carries its own nuisance parameters, which are marginalised over their posterior given the lens data to obtain the lens model posteriors for the observables 
\citep{Pascale_SNH0pe_2025}.}

 {The probability distribution of $H_0$ is obtained by marginalizing over all lens models, weighting each by its ability to reproduce the observed ratios of time delays. For a single lens model $M_\ell$, the posterior on $H_0$ given the light curve data is:}
 {\begin{equation} \label{eq:H0_posterior_single}
    P(H_0 \mid \mathrm{LC},\, M_\ell) 
    = \frac{P(\mathrm{LC} \mid H_0,\, M_\ell) \times P(H_0)}{P(\mathrm{LC} \mid M_\ell)}
\end{equation}}
 {The overall posterior on $H_0$, averaged over $N$ lens models, is then obtained via the law of total probability:
\begin{equation} \label{eq:H0_posterior_avg}
    P(H_0 \mid \mathrm{LC}) 
    = \sum_{\ell=1}^{N} P(H_0 \mid \mathrm{LC},\, M_\ell) \times w_\ell
\end{equation}
where the weights $w_\ell$ for each model (listed in Table~\ref{tab:models}) are 
computed and normalized as:}%
\begin{equation} \label{eq:weights}
    w_\ell \equiv P(M_\ell \mid \mathrm{LC}) 
    = \frac{P(\mathrm{LC} \mid M_\ell) \times P(M_\ell)}
           {\displaystyle\sum_{\ell=1}^{N} P(\mathrm{LC} \mid M_\ell) \times P(M_\ell)}
\end{equation}
 {where $P(\mathrm{LC} \mid M_\ell) = \int P(\mathrm{LC} \mid H_0,\, M_\ell)\,P(H_0)\,dH_0$. 
This assumes partial independence among the lens models and uses only the photometric light curve data. We assume equal prior weights over all lens models, $P(M_\ell) = \mathrm{const}$, such that the weights reduce to $w_\ell \propto P(\mathrm{LC} \mid M_\ell)$ and are determined solely by each model's ability to reproduce the observed data. This analysis uses only the photometric light curve data. Further formalism and methods detailing the $H_0$ inference are given in Appendix~\ref{app:H_0} and in \citet{Pascale_SNH0pe_2025}.}
They also performed a joint analysis including spectroscopic and magnification constraints; however, our analysis uses only weights calculated based on the ability of the models to reproduce the ratios of the photometric time delays (Equation~\ref{eq:weights}), with the values of the weights listed in Table~\ref{tab:models}. The values in Table~\ref{tab:models} are originally presented in Tables~1 and~2 of \citet{Pascale_SNH0pe_2025} and included here for completeness.

 { In addition to the lens model-predicted magnifications, \citet{Pierel_PhotometricTimedelay_2024a} and \citet{chen_SN_2024} measured absolute magnifications for \lensedsn~from the photometry and spectroscopy, respectively. The expected unlensed peak AB magnitude in F277W was measured and averaged for a spectroscopic sample of SN~Ia to get a reference magnitude at $z = 1.78$. F277W was chosen to take advantage of the standardizable nature of SN~Ia in rest frame NIR. The best-fit F277W magnitude of \lensedsn~ from \textsc{BayeSN} (procedure described in Section \ref{sec:obs}) was compared to the reference magnitude get the photometrically measured absolute magnifications. For a detailed explanation, refer Section 4.2 of \citet{Pierel_PhotometricTimedelay_2024a}. The spectroscopic magnifications were estimated similarly, measuring a reference F277W magnitude from a sample of synthetic spectra generated using template spectra and comparing to the observed spectra. Section 7 of \citet{chen_SN_2024} describes the full procedure.}

\section{Modeling the Intrinsic SN~Ia Light Curve}\label{sec: sed modeling}

\subsection{Photometric Corrections} \label{subsec: de-lensing}


In order to test the lens models, the \lensedsn~data had to be corrected to describe the intrinsic \lensedsn~light curve. The hypothesis behind this testing is that if the lens models are correct, removing the magnification from \lensedsn~based on the model predicted values should yield fluxes that are consistent with a typical unlensed SN~Ia at the same redshift (z = 1.78). To further ensure that the data is as close to intrinsic SN~Ia light curves, the light curves for each image are time-shifted based on the inferred photometric time delay to result in a single composite light curve. To create an intrinsic SN light curve, the time delay calculated in \citet{Pierel_PhotometricTimedelay_2024a} for each image was added to the respective epochs for all the three images. Image 2a measured a delay of $-116.6^{+10.8}_{-9.3}$ days while Image 2c measured a delay of $-48.6^{+3.6}_{-4.0}$ days with respect to Image 2b. This time-delay corrected data was then demagnified based on the individual models' predicted magnification for each image (specified in Table \ref{tab:models} to yield a `intrinsic' \lensedsn~light curve. Spectroscopic time delays were also measured for \lensedsn~as stated in \citet{chen_SN_2024} but not used for this analysis. These light curves were analyzed using the SN~Ia SED models described in Section \ref{subsec: sed models}.

\subsection{The SN~Ia SED Models} \label{subsec: sed models}
 {A SN~Ia spectral energy distribution (SED) model is a parametric representation of the time-dependent rest frame SED of SN~Ia based on observed SN populations. This is used to predict the flux of the SN as a function of the wavelength, phase, intrinsic SN properties, distance, etc. Section \ref{subsec: getting mu} provides further detail for how the two SED models described below were used to infer distance modulus.} 

\subsubsection{BayeSN}\label{subsubsec:bayesn}

The first SED model used here is \textsc{BayeSN} \citep{Mandel_2022, Thorp_bayesn, grayling_24_mnras, grayling_bayesn}, a hierarchical Bayesian SED model for SNe~Ia. This framework uses probabilistic generative modeling to account for intrinsic and dust effects underlying the data. The full time- and wavelength-varying SED is described by Equation 1 from \citep{Thorp_bayesn}:
\begin{equation} \label{eqn: bayesn}
    \begin{split}
         - 2.5 \log_{10} [S_s &(t, \lambda_r)/ S_0  (t, \lambda_r)] = \\
         M_0 + &W_0 (t, \lambda_r) + \delta  M^{s} + \theta_1^s W_1 (t, \lambda_r) \\
         + \epsilon^{s} & (t, \lambda_r) + A_{V}^{s} \xi (\lambda_r ; R_{V}^{(s)})
    \end{split}
\end{equation}

where $t$ is the rest-frame phase relative to B-band maximum and $\lambda_r$ is rest-frame wavelength. $S_0(t,\lambda_r)$ is the fixed baseline zeroth-order optical-NIR SN~Ia SED template of \citet{Hsiao_2007} along with the arbitrary normalization constant $M_0$ which is set to -19.5 \footnote{Note that $W_0$ ultimately sets the absolute magnitude of the mean intrinsic SED}. Parameters denoted with $s$ are latent SN parameters and have unique values for each SN $s$. All other parameters are global hyper-parameters shared across the population. They are described in more detail below.

The function, $W_0 (t. \lambda_r)$ manipulates the zeroth-order SED template into a mean intrinsic SED that represents the population. The $W_1 (t, \lambda_r)$ component describes the first mode of the intrinsic SED variation for SNe~Ia. The $\theta_1^s$ coefficient combines with the $W_1 (t, \lambda_r)$ component to represent the `broader-brighter' relation observed in SNe~Ia described in \citet{Phillips_1993}. The relation states that intrinsically brighter light curves evolve over longer timescales around peak brightness. An achromatic, time-independent offset in magnitude is described by $\delta M^s$ for each SN from a normal distribution inferred during model training. The component $\epsilon^{s} (t, \lambda_r)$ accounts for time-varying intrinsic color variations that are not captured by $\theta_1^s W_1 (t, \lambda_r)$. The total V-band extinction is $A_{V}^{s}$ that accounts for one aspect of the host galaxy extinction law. Finally, $R_{V}^{(s)}$ makes up the second part of the host galaxy extinction law and parametrizes the \citet{Fitzpatrick_1999} dust extinction law. 

The model described above gives a rest-frame, host-galaxy dust-extinguished SED model $S_s (t, \lambda_r)$ based on Equation \ref{eqn: bayesn}. This $S_s (t, \lambda_r)$ is scaled based on distance modulus $\mu^s$ before being redshifted and corrected for Milky Way dust extinction to produce an observer-frame SED. This SED is integrated through photometric filters to produce model photometry which is compared with observed photometry to compute a likelihood. For the complete formalism, refer to \citet{Mandel_2022}. 

During training, the model produces posterior estimates for a set of hyperparameters (e.g., spectral weights, dust parameters, and intrinsic scatter), which are marginalized over all latent variables. For simplicity, the posterior means of these hyperparameters are used as point estimates. In distance-fitting mode, the posterior distribution of a supernova’s latent parameters and distance modulus is conditioned on these fixed hyperparameters. This distribution incorporates priors on light-curve shape, color, intrinsic scatter, and host dust. When fitting an individual supernova, the time of maximum light is also treated as a free parameter. By sampling the joint posterior, the marginal posterior distribution of the distance modulus can be approximated. Importantly, this distribution may be asymmetric due to  the non-negativity of the dust effects and is not assumed to be Gaussian.

\subsubsection{SALT3-NIR}\label{subsubsec:salt}

The second SED model used here is \textsc{SALT3-NIR} which is based on a framework that characterizes the spectral flux as a function of wavelength and rest-frame phase. Given below is a brief description of the model based on \citet{SALT3NIR_2022}.  

\begin{equation}
\begin{split}
    F(p, \lambda) =  &x_0 [M_0(p,\lambda) + x_1 M_1(p,\lambda)] \\
    &\cdot \exp (c \cdot C L(\lambda))
\end{split}    
\end{equation}

Three components are determined from the training process characterize the SNe population, with three parameters used to fit each individual SN light curve. The first component, $M_0(p, \lambda)$, represents a baseline SN~Ia; the second, $M_1(p, \lambda)$, provides a first-order linear correction; and the third, $CL(\lambda)$, is a color law accounting for intrinsic color and dust effects based on apparent color. The three fit parameters used for each individual SNe are: $x_0$ (flux normalization), $x_1$ (amplitude of $M_1$, analogous to the stretch parameter $\theta_1^S$ in \textsc{BayeSN}), and $c$ (SN~Ia color).

These three fit parameters are used in the Tripp equation \citep{Tripp_1998},  
\begin{equation}
    \mu = -2.5\log_{10} (x_0) + \alpha x_1 - \beta c - M_0 , 
\end{equation}
to determine the luminosity distances of SNe~Ia. The $\alpha$ and $\beta$ are global parameters that relate luminosity to stretch and color, respectively, while $M_0$ is the absolute magnitude of normal SNe~Ia. Typically, the \textsc{SALT2mu} method described in \citet{Marriner_2011} is used to determine the $\alpha$, $\beta$ and $M_0$ parameters while calculating the distance moduli by minimizing the Hubble residuals in different redshift bins. This leads to a cosmology-independent distance determination. For our analysis, we used $\alpha = 0.14 \pm 0.003$ and $\beta = 3.12 \pm 0.017$ to use the \textsc{SALT2mu} method to determine the distance modulus for each predicted magnification. A key difference to note between \textsc{SALT3-NIR} and \textsc{BayeSN} is how they treat dust during the SED fitting process. The \textsc{SALT3-NIR} has a global dust model that is fit after the light curve of each SN is fit in a two step process. In contrast, \textsc{BayeSN} fits the dust to each SN individually while fitting the light curve allowing for distance inference in a single step. 

\subsection{Extracting Distance Moduli from the Intrinsic Light Curves} \label{subsec: getting mu}
\begin{figure}[t!]
    \centering
    \includegraphics[trim= 0.9cm 1.5cm 1.4cm 1.9cm, clip,width=\linewidth]{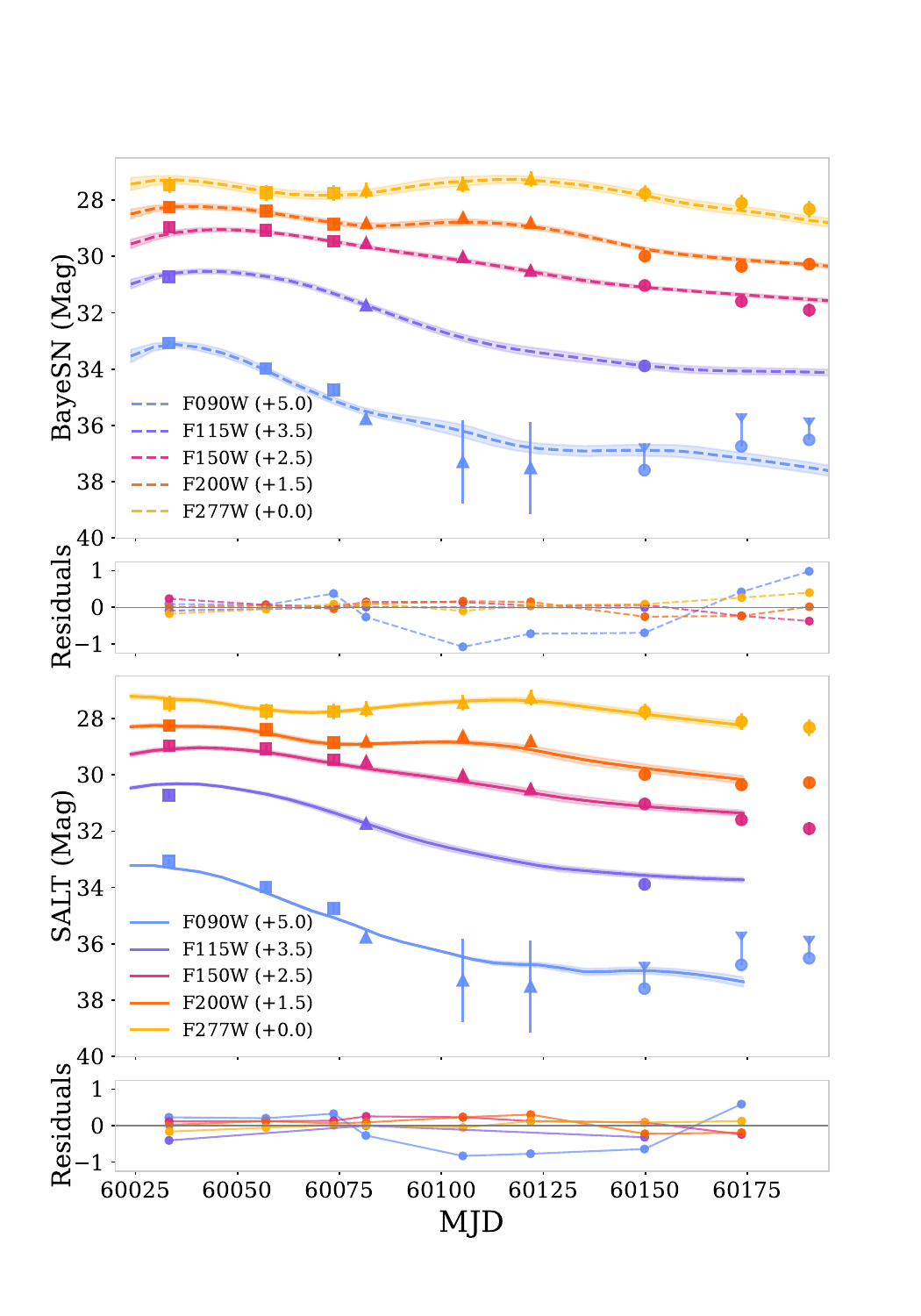}
    \caption{Light curves for \lensedsn~photometry, de-magnified using absolute magnifications as stated in \citet{Pascale_SNH0pe_2025} as predicted by the lens modeling approach of \citet{chen_2020}, are shown fitted with both SED models. The top panel shows \textsc{BayeSN} fit, with the residuals shown in the second panel. The third panel shows the fit using \textsc{SALT3-NIR} with the corresponding residuals in bottom panel. Light curves are color-coded by filter, with vertical offsets (indicated in the legend) applied for clarity. The shaded regions represent the $2 \sigma$ uncertainty intervals for each light curve. Photometric data are overplotted with error bars, also color-coded by filter. Each SN image is represented by a different marker: circles for Image 2a, squares for 2b, and triangles for 2c. }
    \label{fig:lightcurves}
\end{figure}

The `de-magnified' photometry described in Subsection \ref{subsec: de-lensing} was fitted for the distance modulus using the two SN~Ia spectral-energy distribution(SED) models - \textsc{BayeSN} \citep{Mandel_2022} and \textsc{SALT3-NIR} \citet{SALT3NIR_2022} described in Sections \ref{subsubsec:bayesn} and \ref{subsubsec:salt}. \textsc{BayeSN} uses a pre-trained model as a baseline SED to do the fitting.  For our analysis, we used a extended phase NIR model \citep{Grayling_2025} which helped account for the longer rest frame timescale as well as the longer NIR wavelengths of the observations. 
As an example, Figure \ref{fig:lightcurves} shows the fit and residuals of the observed light curves after de-magnification using the lens modeling approach of \citet{chen_2020} with both SN~Ia SED models. We choose the \citet{chen_2020} model because it receives the highest weight in \citet{Pascale_SNH0pe_2025} based on its accuracy in predicting time delays between images. The figure demonstrates that both SED models yield comparable performance, despite differences in their methodology, indicating that the choice of SED model is not a significant source of bias in this analysis.

\begin{figure*}[h]
    \centering
    \includegraphics[trim= 0.2cm 0.2cm 0.2cm 0.2cm, clip,width=0.80\linewidth]{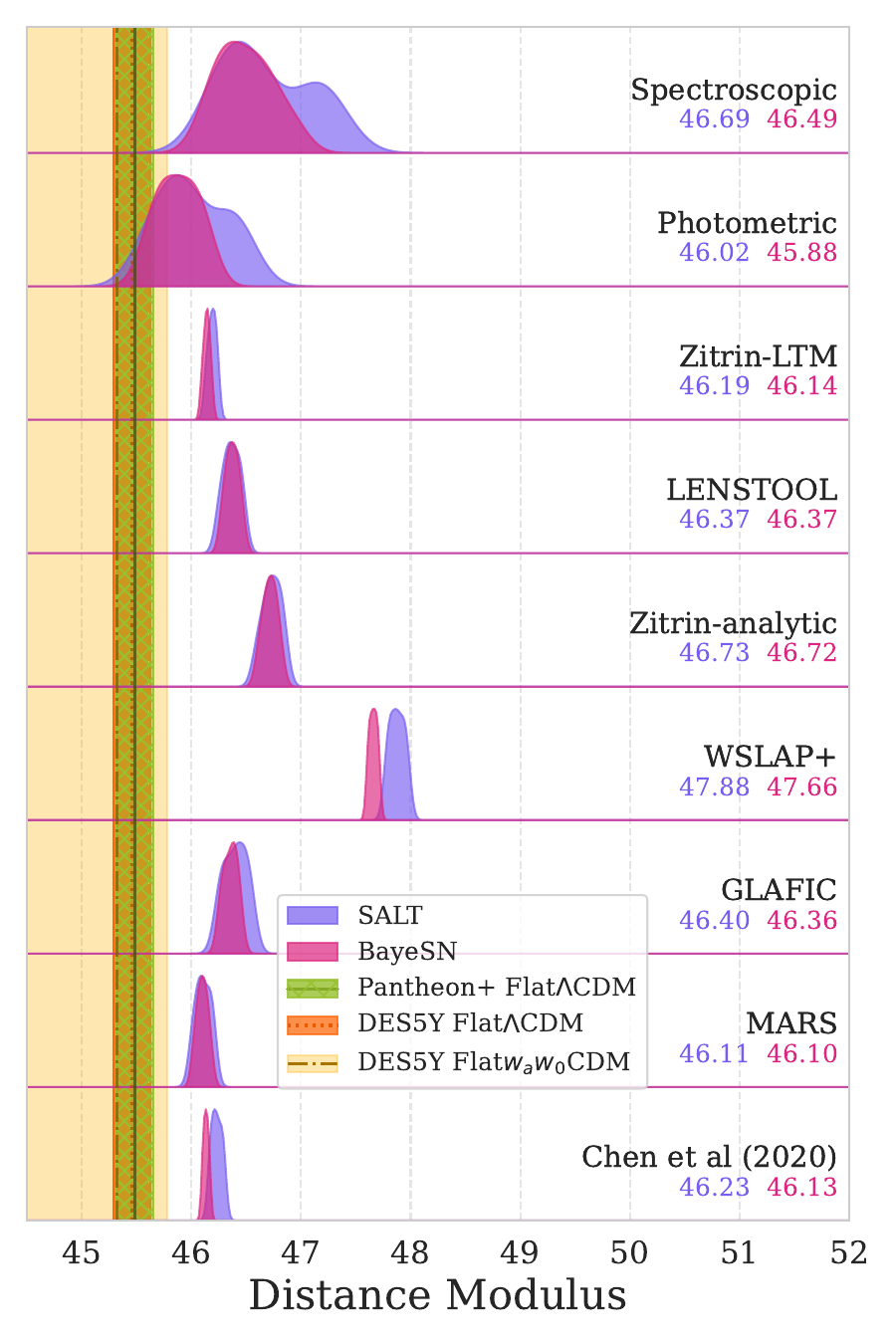}
    \caption{Distance modulus ($\mu$) measurements for each of the lens models described in Section \ref{sec:lens_models} along with the photometric and spectroscopic measurements of magnification from \citep{Pierel_PhotometricTimedelay_2024a} and \citet{Pascale_SNH0pe_2025}. The purple regions show the normalized distribution of distance moduli found using \textsc{SALT3-NIR} while the pink depicts the same found using \textsc{BayeSN}. The expected distance modulus for Flat$\Lambda$CDM (Dotted Orange) and Flat$w_aw_0$CDM (Dot-Dashed yellow) as described in \citet{DES_5Y} at $z=1.783$ are shown as well. Expected distance modulus at $z=1.783$ for Flat$\Lambda$CDM parameters given in \citet{Pantheon_2022} is shown in solid green. Each of the cosmological models has a $5\sigma$ uncertainty region shown in the respective colors as well.} 
    \label{fig:models_error}
\end{figure*}

Each lens model predicts a magnification for each image of \lensedsn~along with the associated uncertainties as specified in Table \ref{tab:models}. For our analysis, the uncertainties here constitute \emph{systematic} shifts in the data rather than random errors. For example, a magnification value higher than the mean for Image 2a would result in all of the photometry for Image 2a shifting together during the demagnification rather than a random scatter about the mean. Thus, these uncertainties in the predicted magnifications cannot be accounted for by simple error propagation. To account for the systematic uncertainties in the predicted magnifications of each image for all the models, the photometry was de-magnified using a range of predicted magnifications for each image (specified by the value itself, and the upper and lower $1 \sigma$ limit of expected magnification) for every lens model. Each iteration of the modified image was then fitted with both SN~Ia SED models to ensure that the full parameter space was explored. The resulting distribution of distance modulus values are shown in Figure \ref{fig:models_error}. This procedure provides the most conservative and rigorous bound on the systematic impact of an error in the magnification. 

Finally, to test the lens model predictions, the `de-magnified' distance moduli were compared to the expected distance moduli for the different inferred cosmologies at $z=1.783$. We use cosmological parameters inferred by the \citet{Pantheon_2022} $\text{Flat}\Lambda\text{CDM}$ as well as \citet{DES_5Y} $\text{Flat}\Lambda\text{CDM}$ and $\text{Flat}w_aw_0\text{CDM}$ shown in Figure \ref{fig:models_error}. All of the included cosmological parameters were inferred using SN only and were not combined with any other probes. This ensures that the $3\sigma$ error regions given for each cosmology includes the intrinsic SN scatter. 

The final aspect to consider is the method used by the SN~Ia fitters to obtain a distance modulus. This analysis doesn't use any absolute calibrators (eg: Cephieds etc), without which SN~Ia SED model fitters can only provide relative distance estimates. They adopt a fiducial value of $H_0$ to put photometric distance estimates on a distance scale. Thus, for a self-consistent analysis, the distance can and must be trivially rescaled to ensure that a single $H_0$ value is being used for both the distance estimates and the cosmological model. \textsc{BayeSN} uses an assumed value of $H_0 = 73.24$ while \textsc{SALT3-NIR} uses the $H_0$ value from the Pantheon+ cosmological constraints \citep{Pantheon_2022}. 
In order to compare the extracted distance moduli with the cosmological constraints from Pantheon+ and DES 5Y survey results, they need to be rescaled to the same $H_0$ value, else the differing values cause a persistent shift in the inferred distance moduli. For our analysis we use the Pantheon+ value of $H_0 = 73.6 \pm 1.1$ \citep{Pantheon_2022} for all the cosmological constraints, with the rescaling based on the ratio of the new $H_0$ value to the old assumed $H_0$.

\section{Testing Lens Model Systematics} \label{sec:results}
We analyzed the predicted magnifications of all seven lens models used in \citep{Pascale_SNH0pe_2025} as well as the magnifications measured directly from the photometric and spectroscopic data. 
Using only the lens model–predicted magnifications, we find weighted mean distance moduli of $46.35^{+0.47}_{-0.32}$ with \textsc{SALT3-NIR} and $46.33^{+0.49}_{-0.26}$ with \textsc{BayeSN}. These values were computed using the TD-only weights from \citet{Pascale_SNH0pe_2025}. We emphasize that these weighted mean values only include the lens models from Section \ref{sec:lens_models}, and explicitly do not include the distance moduli extracted from the magnifications directly measured from photometry and spectroscopy, which are reported separately in Table~\ref{tab:models}.

\begin{figure}[t]
    \centering
    \includegraphics[trim= 0.3cm 0.3cm 0.3cm 0.3cm, clip,width=\linewidth]{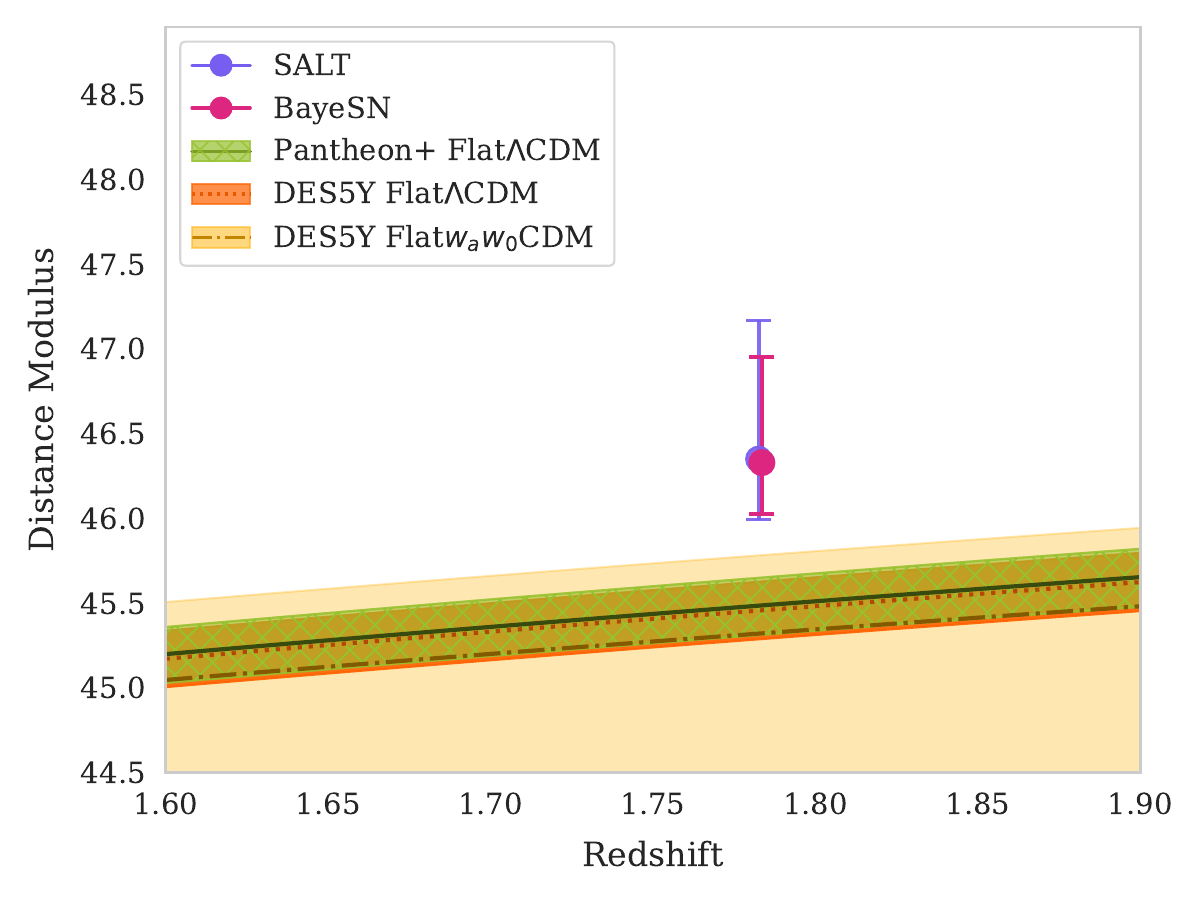}
    \caption{The weighted distance modulus for \textsc{SALT3-NIR} (purple) vs \textsc{BayeSN} (pink). The error bars for the two points show the $5~\sigma$ spread for the distribution. The distance modulus - redshift relation for three cosmological constraints are plotted as well. The DES5Y Flat$\Lambda$CDM is depicted in a orange dotted line. The DES5Y Flat$w_0w_a$CDM is shown in dotted-dashed yellow. The green solid line is the Pantheon+ Flat$\Lambda$CDM cosmology. Each cosmological model is accompanied by $5\sigma$ error regions in the respective colored shaded regions. }
    \label{fig:SVB_LCDM}
\end{figure}

Figure \ref{fig:SVB_LCDM} shows that the weighted mean distance modulus is systematically higher than expected from SN~Ia–based cosmological constraints, indicating that the lens models for \lensedsn~over-predict magnifications. This trend persists regardless of which SN~Ia SED model is used to fit the light curves. As shown in Figure \ref{fig:lightcurves}, both SED models yield similar residuals, demonstrating that the over-prediction does not arise from the light curve fitting itself. To place these results in context, we compare the lens model–predicted distance moduli with the values expected at $z=1.783$ under Flat$\Lambda$CDM and Flat$w_aw_0$CDM cosmologies. Specifically, we adopt the Flat$\Lambda$CDM model from \citet{Pantheon_2022} (hereafter Pantheon+ Flat$\Lambda$CDM) and both Flat$\Lambda$CDM and Flat$w_aw_0$CDM from \citet{DES_5Y} (hereafter DES5Y Flat$\Lambda$CDM and DES5Y Flat$w_aw_0$CDM). Since these cosmological constraints are derived solely from SN datasets, the $5\sigma$ uncertainty regions in Figure \ref{fig:SVB_LCDM} naturally encompass the intrinsic scatter of SNe~Ia.

Figure \ref{fig:models_error} shows the probability density functions (PDF) of the distance moduli at $z = 1.783$ for each lens model, as well as the photometric and spectroscopic analyses, for both SED models. These PDFs incorporate the systematic uncertainties associated with the predicted magnifications of each of the three Images of \lensedsn~(2a, 2b, 2c) for every individual lens model. \textsc{SALT3-NIR} values are shown in purple while \textsc{BayeSN} values are in pink. The orange dotted line marks the distance modulus predicted by the DES5Y Flat$\Lambda$CDM model \citep{DES_5Y} at $z=1.783$, with its $5\sigma$ uncertainty shown by the orange shaded band. The yellow dash-dotted line indicates the DES5Y Flat$w_aw_0$CDM prediction \citep{DES_5Y}, with the corresponding $5\sigma$ region shaded in yellow. The solid green line shows the Pantheon+ Flat$\Lambda$CDM prediction \citep{Pantheon_2022} at the same redshift, with the green shaded region highlighting the $5\sigma$ uncertainty.

It is evident from Figure~\ref{fig:models_error} that the lens model–predicted magnifications (as seen in Table \ref{tab:models}) are systematically biased high when compared to the predictions of both the Pantheon+ and DES-5Y cosmological models. In contrast, the photometric estimate of $\mu$ agrees well with these fiducial cosmologies, independent of the SED model used. The spectroscopically derived $\mu$ also shows tension with the fiducial cosmologies, but its much larger uncertainty compared to the lens models reduces the significance of this offset. A similar trend is suggested in Figure~3 of \citet{Pascale_SNH0pe_2025}, though it is not a central focus of that analysis. Here, we explicitly highlight and confirm this behavior through our independent methodology.

To quantify the offset, we define $\Delta \mu$ as the difference between the SN-inferred and cosmology-predicted distance modulus: $\Delta\mu \equiv \mu_{\rm SN} - \mu_{\Lambda{\rm CDM}}$, measured at $z=1.783$, across the seven lens model for both SED models and three cosmological baselines. For \textsc{SALT}, the grand (``pooled'') mean $\Delta\mu$ is $1.073 \pm 0.085$~mag under Pantheon+ flat $\Lambda$CDM, $1.102 \pm 0.082$~mag under DES5Y flat $\Lambda$CDM, and $1.235 \pm 0.149$~mag under DES5Y $w_0w_a$CDM. For \textsc{BayeSN}, we find similarly elevated values of $1.037 \pm 0.071$~mag (Pantheon+), $1.066 \pm 0.067$~mag (DES5Y Flat$\Lambda$CDM), and $1.199 \pm 0.141$~mag (DES5Y Flat$w_0w_a$CDM). 

These grand means reflect consistent overestimation of the lensing magnification relative to the cosmological expectation, regardless of SED model. However, there is also substantial variation between individual lens models. For \textsc{SALT}, the per-model mean $\Delta\mu$ values span $0.622$–$2.391$~mag (Pantheon+), $0.652$–$2.421$~mag (DES5Y Flat$\Lambda$CDM), and $0.784$–$2.553$~mag (DES5Y $w_0w_a$CDM). For \textsc{BayeSN}, the corresponding ranges are $0.640$–$2.199$~mag (Pantheon+), $0.669$–$2.229$~mag (DES5Y Flat$\Lambda$CDM), and $0.802$–$2.361$~mag (DES5Y $w_0w_a$CDM). These wide ranges suggest that while the overall bias is robust, it is not uniform across all lens model predictions.

In all cases, the quoted ``$\pm$'' values denote the pooled per-point $1\sigma$ dispersion, calculated across all lens model data points per cosmology and SED model, and should not be interpreted as the uncertainty on the mean. For comparison, the uncertainty on each per-model mean—combining both photometric scatter and the uncertainty in the reference cosmological distance modulus—is typically $\sim0.039$–$0.042$~mag for Pantheon+ Flat$\Lambda$CDM, $\sim0.034$–$0.039$~mag for DES5Y Flat$\Lambda$CDM, and $\sim0.094$–$0.096$~mag for DES5Y Flat$w_0w_a$CDM, for both \textsc{SALT} and \textsc{BayeSN}.

Figure \ref{fig:models_error} also illustrates slight differences in distance modulus estimates between \textsc{SALT3-NIR} and \textsc{BayeSN}, which arise from differences in how each SED model treats different aspects during fitting. In \textsc{SALT3-NIR}, uncertainties are evaluated in two stages: first in fitting the light-curve parameters, and second in deriving the distance modulus with the \textsc{SALT2mu} method. By contrast, \textsc{BayeSN} infers the distance modulus in a single step that jointly accounts for correlations among light-curve shape, color, and luminosity distance. This contrast is further illustrated in Figure \ref{fig:SVB_LCDM}, which compares the weighted average $\mu$ values from each SED model. The \textsc{SALT3-NIR} estimate (purple) lies in slightly lower tension with the cosmological baselines than the \textsc{BayeSN} estimate (pink), though the difference is not statistically significant.

Another factor that can influence the distance inference is biases introduced in the light curve fitting. Selection effects and inference priors in SN Ia light curve fitting can introduce systematic biases in distance modulus of order $ \sim 0.05 - 0.10$ mag. While we do not perform detailed simulations to correct for these biases in this work, we qualitatively assess their likely direction based on previous studies. For \textsc{SALT3-NIR} and \textsc{BayeSN}, a negative bias is expected, particularly at higher redshift or lower signal-to-noise, where selection effects favor intrinsically brighter SNe Ia, leading to underestimated distances \citet{Pantheon_2022, Hounsell_2018}. Nevertheless, given that the cosmologies used in this comparison were derived using the same SED models applied to large samples of Type Ia supernovae from DES5Y and the Pantheon analysis, we should expect that the Type Ia SN H0pe would be consistent with the same fiducial cosmologies if the lens models describe the system accurately. 

\begin{figure*}[t]
    \centering
    \includegraphics[trim= 0.0cm 0.0cm 0.0cm 0.0cm, clip,width=0.8\linewidth]{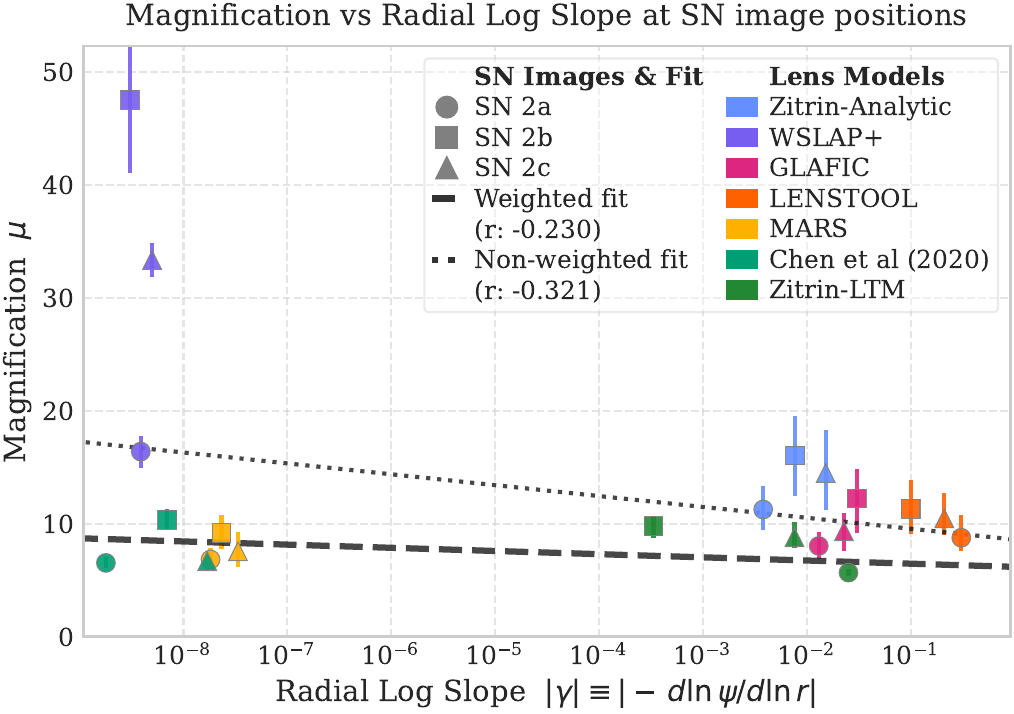}
    \caption{Magnification ($\mu$) of each SN image 2a, 2b, 2c (circle, square, triangle) for all the seven lens model vs the local radial logarithmic slope of the lensing potential $|\gamma| = |- d \ln \psi / d \ln r|$ at those positions. Each marker is colored according to the lens models. The error bars are $2 \sigma$ uncertainty regions as reported in \citep{Pascale_SNH0pe_2025}. The dashed line shows a weighted linear fit in $\mu$ versus $\gamma$ (r = -0.230) while the dotted line shows the unweighted linear fit (r=-0.321). The negative correlation here suggests that a steeper slope leads to a lower magnification value. }
    \label{fig:Lens_slopes}
\end{figure*}

 {Figure \ref{fig:Lens_slopes} shows the magnification of each SN image as a function of the local radial logarithmic slope ($\gamma$) of the projected lensing potential ($\psi$) at the SN image locations for each lens model. The radial log slope is calculated as:}
\begin{equation}
    |\gamma| = |- d\ln \psi/d \ln r| 
\end{equation}
The figure also shows the weighted and unweighted linear fits of the magnification as a function of the radial log slope, with Pearson $r = -0.230$ (weighted) and $r = -0.321$ (unweighted). Weights used for the fit were the inverse square of the magnification errors reported for each model. Based on the negative correlation above,  we can speculate that the true slope of the potential may be steeper than the one from current lens models.

\section{Discussion and Conclusion} \label{sec: concl}

Strongly-lensed transients are a new avenue to understand many different aspects of astrophysics - dark matter, dark energy, high redshift SNe, and more. To leverage the full potential of these unique systems, analyses must be robust and lens models must be well-tested. Here, we test the robustness of the lens models used in the cosmological analysis of \lensedsn.

We de-magnified the \lensedsn~photometry using the absolute magnifications predicted by each of the lens models. We combined this with the time-delays measured in \citet{Pierel_PhotometricTimedelay_2024a} to produce an ``intrinsic'' SN light curve from the de-magnified photometry. Next, we used existing Type Ia SED models to estimate the distance modulus ($\mu$) for each predicted magnification. Finally, we compared these $\mu$ values with those expected from $\Lambda$CDM to assess the performance of the lens models. Our analysis shows that the calculated distance moduli are systematically higher than expected at $z=1.783$ as seen in Figure \ref{fig:models_error}, suggesting that the lens models overestimate the absolute magnifications at the $\sim 4.5\sigma$ level.

For each of the seven lens models as well as the spectroscopic and photometric methods of measuring magnifications, the derived distance moduli are greater than the value expected for each of the cosmological models at $z=1.783$. To quantify this offset, we define $\Delta \mu \equiv \mu_{\text{SN}} - \mu_{\Lambda\text{CDM}}$, the difference between the SN-inferred and cosmology-predicted distance moduli. Across all lens models, SED models and cosmological baselines, we consistently find $\Delta \mu > 1$~mag. For \textsc{SALT}, the pooled mean offset is $1.073 \pm 0.085$~mag under the Pantheon+ flat $\Lambda$CDM, $1.102 \pm 0.082$~mag under DES5Y flat $\Lambda$CDM, and $1.235\pm0.149$~mag under DES5Y $w_0w_a$CDM. Similarly, for \textsc{BayeSN}, we find offsets of $1.037 \pm 0.071$, $1.066 \pm 0.067$ and $1.199 \pm 0.141$~mag under the same three cosmologies, respectively. These results suggest that current lens-model magnifications may be systematically overestimated, independent of the assumed SN~Ia SED model or cosmological baseline. Rather than undermining their utility, this highlights how cluster-lensed SNe provide a rare and powerful testbed to identify and correct systematics in lens modeling.

Although the degree of offset varies across models, all predict values higher than expected at this redshift. This is surprising because the magnifications of each model were determined in a blinded analysis (as specified in \citet{Pascale_SNH0pe_2025}) and used a variety of techniques to construct the lens models. Considering the different methods combined with the blinded analysis, some scatter about the true value is normal and would be expected. However, none of the models/methods lead to an underestimation of the distance modulus. This systematic shift toward higher distance modulus values suggests an underlying cause beyond random scatter. We also measured the logarithmic slopes of the lensing potential for each lens model at the locations of the \lensedsn~images. Figure \ref{fig:Lens_slopes} shows that shallower models predict higher magnifications while steeper models predict lower magnifications. This inverse correlation is modest but consistent, with a Pearson correlation coefficient of $r = -0.230$ when weighting by the uncertainties in magnification, and $r = -0.321$ when unweighted. This trend suggests that the true potential of the G165 system may be steeper than predicted by any current lens models, and potentially contributing to the systematic bias seen in Figure \ref{fig:models_error}.

The magnification overestimates were first seen in Figure 3 of \citet{Pascale_SNH0pe_2025}. Although the overestimates were recognized, they could not be corrected due to the strictly enforced post-unblinding rules agreed upon beforehand \citep{Pascale_SNH0pe_2025}. Revised lens models will reduce this bias (Pascale et al. 2026, \textit{in prep}), which could be indicative of a systematic error that we have been unaware of due to the lack of avenues for testing these lens models. One possible explanation is that the discrepancy arises from the cosmological models themselves. However, the cosmological models have held up well against various testing methods until now and been proven to match observations well. In contrast, there have been few avenues to test cluster scale lens models until now. This system is notable in that it includes a transient and multiple image systems at different spectroscopically confirmed redshifts—features that help break the mass sheet degeneracy and enable the construction of more robust lens models. We refer the reader to \citet{Pascale_SNH0pe_2025} for a more thorough discussion on the impact of mass-sheet and more generally mass-slope degeneracy in this field. This does not mean that the $\Lambda \text{CDM}$ cosmologies are infallible, but merely that we must conduct our due diligence and ensure that other sources of error have been accounted for. 

Another avenue that could be affecting this analysis is the photometry. \lensedsn~ did not have a template image to get precise SN flux measurements at the time of this analysis and a lens model was used to subtract the host flux. Our initial calculations suggest that the improvement expected in photometric data from the inclusion of a template will not be enough to account for the overestimated distance moduli, though it may be a contributing factor. We plan to address this in our future work, now that a template image has been obtained in a JWST Cycle 3 program (GO 4744, PI Frye \& Pierel). 

An important next step is to consider how these results impact the broader goal of using glSNe to measure $H_0$. Assuming fixed time delays, modifying the predicted absolute magnifications to bring the lens model–based distance modulus into agreement with current cosmological constraints would result in a higher inferred $H_0$. In the case of \lensedsn, such a revision could shift the inferred value further from the CMB-based measurement, potentially altering the degree of Hubble tension--at least for this individual object. However, it is important to note that magnification predictions are inversely correlated with time-delay predictions. If the lens models are improved to have lower magnifications to resolve the overestimation, we would be sampling larger time delays when measuring $H_0$. This is seen in the $H_0$ inference done in \cite{Pascale_SNH0pe_2025}, where the TD-only weighted $H_0$ value is lower than those weighted using magnifications. It is important to note that these findings are based on the currently available lens models and photometric data. Both are expected to improve, with new lens model reconstructions currently underway (Pascale et al. 2026, \textit{in prep.}) and updated measurements of the SN photometry and time delays made possible by a recently obtained template image (Agrawal et al. 2026, \textit{in prep}). These updates will provide a more robust basis for interpreting the magnification predictions and their cosmological implications of \lensedsn.

There are several methods to mitigate lens modeling systematics in $H_0$ inference. \cite{Liu_Oguri_2025} demonstrates that additional lensing evidence such as additional multiply lensed systems/events leads to improvements in lens modeling. The paper also highlights the correlation between the Hubble constant and magnification, which motivates implementing a magnification-weighting scheme—that is, properly sampling the posterior (e.g., \cite{Pascale_SNH0pe_2025}). Another option is to use the magnifications as a constraint in the lens modeling itself, as was done for the unblinded models in \cite{Pascale_SNH0pe_2025}. 

Cluster-scale lens models are inherently complex, built on current understanding of dark matter and gravity together with simplifying assumptions. Until now, there have been limited opportunities to test these methods. Our analysis shows that lensed supernovae are useful probes of lensing at these scales. It is yet unclear if this overestimation is due to an underlying assumption made when building the lens model or some new underlying physics. Both possibilities present exciting avenues forward and may be key to improving our models to leverage them for time delay cosmography. It is clear that a larger sample of cluster-lensed SNe~Ia is needed to disentangle the true underlying cause and will lead to a substantial improvement in our lens modeling. Our analysis also underscores the importance of using magnification constraints in lens modeling for time delay cosmography as well as using multiple lens models for $H_0$ inference. Figure \ref{fig:models_error} highlights how reliance on a single lens model, without independent avenues for testing, can allow hidden systematics to propagate into the final result. Improving these lens models is a crucial step towards high-precision cosmology with lensed phenomena. \lensedsn~highlights the power of cluster-lensed SNe to reveal systematics and improve the robustness of $H_0$ inference. 

\section{Future Work} \label{sec:future work}
We are investigating why all lens models consistently over predict the magnifications. Our first step is to repeat the analysis using updated SN photometry corrected for host flux contamination, thanks to a new template image obtained in a JWST Cycle 3 program. Next, we will probe the discrepancy between lens-model predictions and cosmological values at $z=1.783$, to determine whether it stems from modeling issues or potentially points to new physics. We are also applying this method to other lensed SNe, particularly SN Requiem and SN Encore—two events in the same source-lens system expected to provide tighter constraints—and will extend the analysis to future strongly lensed supernovae.

\section{Acknowledgments} \label{sec:Acknow}
This paper is based in part on observations with the NASA/ESA Hubble Space Telescope obtained from the Mikulski Archive for Space Telescopes at STScI. We thank the DDT and JWST scheduling team at STScI for extraordinary effort in getting the DDT observations used here scheduled quickly. AA acknowledges support from AST-2206195 (P.I. Narayan), to develop anomaly detection methods to identify lensed supernovae, and HST-GO-17128 (PI: R. Foley) to adapt BayeSN to model \emph{JWST} and \emph{HST} observations of type Ia supernovae. AA gratefully acknowledge support from NSF AST-2421845 and support from the Simons Foundation as part of the NSF-Simons SkAI Institute as a 2026 SkAI Graduate Fellow at UIUC. JDRP is supported by NASA through a Einstein Fellowship grant No. HF2-51541.001 awarded by the Space Telescope Science Institute (STScI), which is operated by the Association of Universities for Research in Astronomy, Inc., for NASA, under contract NAS5-26555. GN gratefully acknowledges NSF support from NSF CAREER grant AST-2239364, supported in-part by funding from Charles Simonyi, to model type Ia supernovae with ground- and space-based data, AST 2206195 to develop anomaly detection methods to identify lensed supernovae, aand DOE support through the Department of Physics at the University of Illinois, Urbana-Champaign (Grant No. 13771275) to deploy the lensed SN modeling pipeline from this work for the Vera C. Rubin Observatory.  GN also gratefully acknowledge support from NSF AST-2421845 and support from the Simons Foundation as part of the NSF-Simons SkAI Institute, and NSF OAC-1841625, OAC-1934752, OAC-2311355, AST-2432428 as part of the Scalable Cyberinfrastructure for Multi-messenger Astrophysics (SCIMMA) team. MG and KSM are supported by the European Union’s Horizon 2020 research and innovation programme under European Research Council Grant Agreement No 101002652 (BayeSN; PI K. Mandel) and Marie Skłodowska-Curie Grant Agreement No 873089 (ASTROSTAT-II). This material is based upon work supported by the National Science Foundation Graduate Research Fellowship Program under Grant No. DGE 21-46756. RAW acknowledges support from NASA JWST Interdisciplinary Scientist grants NAG5-12460, NNX14AN10G and 80NSSC18K0200 from GSFC. 
\bibliography{references}{}
\bibliographystyle{aasjournal}



\appendix
\vspace{-2em}
\section{BayeSN Fits and Corner Plots}
\label{app:bayesn_corner}

\begin{figure*}[h]
    \centering
    \includegraphics[trim= 0.9cm 0.9cm 0.9cm 0.9cm, clip,width=0.75\linewidth]{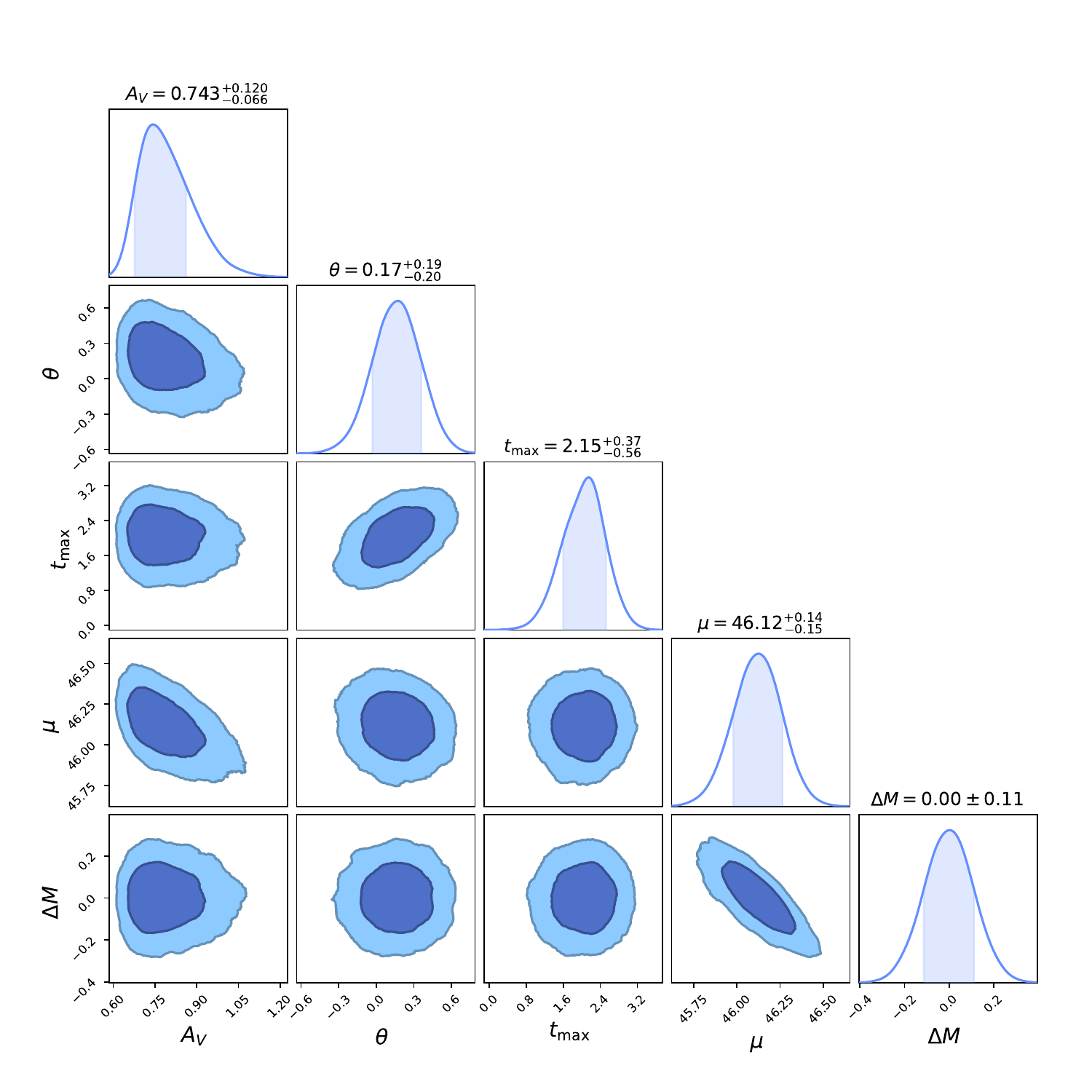}
    \caption{Corner plot for the posterior distributions of the \textsc{BayeSN} fit parameters using the mean magnifications from the \citet{chen_2020} lens model. The contours show the $68\%$ and $95\%$ credible intervals, while the diagonal panels display the marginalized distributions for $A_V$, $\theta$, $t_{\max}$, $\mu$ and $\Delta M$.}
    \label{fig:corner}
\end{figure*}

The \textsc{BayeSN} model described in Section \ref{subsubsec:bayesn} were used to fit the de-lensed light curves for each of the seven independent cluster lens models, as well as for both the photometrically and spectroscopically measured magnifications. This ensures that the inference of the intrinsic SN~Ia parameters is consistently propagated across the full range of lensing scenarios explored in this work. Figure~\ref{fig:corner} shows the corner plot of the posterior distributions for the key \textsc{BayeSN} light-curve parameters obtained when adopting the mean magnifications of the three SN images predicted by the \citet{chen_2020} lens model. The fitted parameters include the host-galaxy dust extinction $A_V$, the light-curve shape $\theta$, the time of maximum $t_{\max}$, the dstance modulus $\mu$ and the intrinsic luminosity scatter $\Delta M$. The two-dimensional contours highlight the covariance between these quantities, while the marginalized histograms along the diagonal display the recovered uncertainties. As seen in Figure~\ref{fig:corner}, the posteriors are unimodal and well-sampled, with clear $68\%$ and $95\%$ credible intervals and no evidence for unconstrained directions in parameter space. The inferred parameter values are consistent with expectations from the \textsc{BayeSN} training set and remain stable when alternative lens models are used.

\section{Formalism for the $H_0$ posterior distribution}\label{app:H_0}
Here, we present a self-contained derivation of the Bayesian formalism for inferring $H_0$ from multiply-imaged supernova light curves , following the frameworks of \citet{Kelly_2023_b} and \citet{Pascale_SNH0pe_2025}. Before any lens model or H$_0$ analysis, we fit the LCs of the multiple images to obtain samples from the posterior:
\begin{equation}
    P(\theta, \nu \mid \mathrm{LC}) = \frac{P(\mathrm{LC} \mid \theta, \nu)\,\tilde{P}(\theta)\,P(\nu)}{\tilde{P}(\mathrm{LC})}
    \label{eq:joint_posterior}
\end{equation}
where $\theta$ = lensing observables (latent variables), e.g.\ time delay, and $\nu$ = SN-model-specific parameters, e.g.\ light-curve shape, etc. $P(\nu)$ is the prior on SN-parameters, and $\tilde{P}(\theta)$ is an interim prior on lensing observables (not to be confused with lens model predictions $P(\theta \mid H_0, M_\ell)$). We assume $\tilde{P}(\theta)$ is uniform over a region of $\theta$-space, $\Omega$, that is much wider than the support of the likelihood $P(\mathrm{LC} \mid \theta, \nu)$ (i.e.\ $\{\theta : P(\mathrm{LC} \mid \theta, \nu) > 0\}$):
\begin{equation*}
    \tilde{P}(\theta) =
    \begin{cases}
        \tilde{V}^{-1}, & \theta \in \Omega \\
        0,              & \text{otherwise}
    \end{cases}
    \quad \text{where } \tilde{V} = \int_\Omega d\theta
\end{equation*}
and
\begin{align*}
    \tilde{P}(\mathrm{LC}) &\equiv \int P(\mathrm{LC} \mid \theta, \nu)\,\tilde{P}(\theta)\,P(\nu)\,d\theta\,d\nu \\
                            &= \tilde{V}^{-1} \int P(\mathrm{LC} \mid \theta, \nu)\,P(\nu)\,d\theta\,d\nu.
\end{align*}

Now, the samples from the posterior \eqref{eq:joint_posterior} are used to construct a numerical approximation (GMM) of the marginal posterior:
\begin{align}
    P(\theta \mid \mathrm{LC}) &= \int P(\theta, \nu \mid \mathrm{LC})\,d\nu \notag \\
                                &= \frac{\tilde{P}(\theta)}{\tilde{P}(\mathrm{LC})} \int P(\mathrm{LC} \mid \theta, \nu)\,P(\nu)\,d\nu.
    \label{eq:marginal_posterior}
\end{align}
This can be rewritten as:
\begin{equation}
    \int P(\mathrm{LC} \mid \theta, \nu)\,P(\nu)\,d\nu = \frac{\tilde{P}(\mathrm{LC})}{\tilde{P}(\theta)}\,P(\theta \mid \mathrm{LC}).
    \label{eq:rewrite}
\end{equation}

\hrule\vspace{6pt}

To perform the $H_0$ analysis with lens model $M_\ell$, we need to compute $P(\mathrm{LC} \mid H_0, M_\ell)$ up to an unknown constant that is independent of $H_0$ and $M_\ell$:
\begin{align}
    P(\mathrm{LC} \mid H_0, M_\ell)
        &= \iint P(\mathrm{LC} \mid \theta, \nu)\,P(\nu)\,P(\theta \mid H_0, M_\ell)\,d\nu\,d\theta \notag \\
        &= \int \left[\int P(\mathrm{LC} \mid \theta, \nu)\,P(\nu)\,d\nu\right] P(\theta \mid H_0, M_\ell)\,d\theta \notag \\
        &= \tilde{P}(\mathrm{LC}) \int \frac{P(\theta \mid \mathrm{LC})}{\tilde{P}(\theta)}\,P(\theta \mid H_0, M_\ell)\,d\theta \notag \\
        &= \tilde{V}\,\tilde{P}(\mathrm{LC}) \int_\Omega P(\theta \mid \mathrm{LC})\,P(\theta \mid H_0, M_\ell)\,d\theta \notag \\
        &= \mathrm{const} \times \int_\Omega P(\theta \mid \mathrm{LC})\,P(\theta \mid H_0, M_\ell)\,d\theta,
    \label{eq:LC_H0_Me}
\end{align}
where $\mathrm{const} = \int P(\mathrm{LC} \mid \theta, \nu)\,P(\nu)\,d\theta\,d\nu$, which is independent of $H_0$ and $M_\ell$ and will cancel in subsequent calculations (so we do not need to evaluate it).

As shown in Equation~\ref{eq:LC_H0_Me}, the lens model prediction for the lensing observables enters through $P(\theta \mid H_0, M_\ell)$. Each lens model carries its 
own set of nuisance parameters $\phi_\ell$ (e.g.\ describing the dark matter distribution) that are independent of the LC fitting. Analogously to the SN fitting, these nuisance parameters are marginalised over their posterior given the lens data to obtain the lens model posteriors for the observables. Since $\phi_\ell$ is independent of the LC inference, this marginalisation does not affect the derivation that follows.

Now we compute the posterior of $H_0$ for one lens model $M_\ell$.

 {Assumption:} The prior $P(H_0, M_\ell) = P(H_0)\,P(M_\ell)$, i.e.\ $H_0$ and $M_\ell$ have independent priors, so $P(H_0 \mid M_\ell) = P(H_0)$.

By Bayes' theorem:
\begin{equation}
    P(H_0 \mid \mathrm{LC}, M_\ell) = \frac{P(\mathrm{LC} \mid H_0, M_\ell) \times P(H_0)}{P(\mathrm{LC} \mid M_\ell)}.
    \label{eq:H0_posterior_single}
\end{equation}

Using the above, we compute the posterior of $H_0$, averaging over $N$ Lens Models. 

By the law of total probability:
\begin{align}
    P(H_0 \mid \mathrm{LC})
        &= \sum_{\ell=1}^{N} P(H_0 \mid \mathrm{LC}, M_\ell) \times P(M_\ell \mid \mathrm{LC}) \notag \\
        &= \sum_{\ell=1}^{N} P(H_0 \mid \mathrm{LC}, M_\ell) \times w_\ell,
    \label{eq:H0_posterior_avg}
\end{align}
where the weights are
\begin{equation}
    w_\ell \equiv P(M_\ell \mid \mathrm{LC}) = \frac{P(\mathrm{LC} \mid M_\ell) \times P(M_\ell)}{\displaystyle\sum_{\ell=1}^{N} P(\mathrm{LC} \mid M_\ell) \times P(M_\ell)},
    \label{eq:weights}
\end{equation}
where
\begin{equation*}
    P(\mathrm{LC} \mid M_\ell) = \int P(\mathrm{LC} \mid H_0, M_\ell) \times P(H_0)\,dH_0.
\end{equation*}

We assume equal prior weights over all $N$ lens models, 
$P(M_\ell) = \mathrm{const}$, so that Equation~\ref{eq:weights} simplifies to
\begin{equation*}
    w_\ell = \frac{P(\mathrm{LC} \mid M_\ell)}{\displaystyle\sum_{\ell=1}^{N} 
    P(\mathrm{LC} \mid M_\ell)}.
\end{equation*}
This assumption is not always stated explicitly in the literature 
\citep[e.g.][]{Pascale_SNH0pe_2025}, but is required for the prior weights to 
drop out. Unequal prior weights $P(M_\ell)$ could in principle be used to encode 
information about how well each lens model reproduces other lensing observables 
beyond the glSN light curve data.

\end{document}